\documentclass[12pt,preprint]{aastex}

\shorttitle{TWO-CURRENT-SHEET RECONNECTION MODEL}
\shortauthors{Zhang et al.}

\begin{document}

\title{TWO-CURRENT-SHEET RECONNECTION MODEL OF INTERDEPENDENT FLARE
AND CORONAL MASS EJECTION}

\author{Y. Z. ZHANG\altaffilmark{1}, J. X. WANG\altaffilmark{1} AND  Y. Q. HU\altaffilmark{2}
}

\altaffiltext{1}{National Astronomical Observatories, Chinese
Academy of Sciences, Beijing 100012, China}
\altaffiltext{2}{School of Earth and Space Sciences, University of
Science and Technology of China, Hefei 230026, China}

\begin{abstract}
Time-dependent resistive magnetohydrodynamic simulations are
carried out to study a flux rope eruption caused by magnetic
reconnection with implication in coexistent flare-CME (coronal
mass ejection) events. An early result obtained in a recent
analysis of double catastrophe of a flux rope system is used as
the initial condition, in which an isolated flux rope coexists
with two current sheets: a vertical one below and a transverse one
above the flux rope. The flux rope erupts when reconnection takes
place in the current sheets, and the flux rope dynamics depends on
the reconnection sequence in the two current sheets. Three cases
are discussed: reconnection occurs (1) simultaneously in the two
current sheets, (2) first in the transverse one and then in the
vertical, and (3) in an order opposite to case 2. Such a
two-current-sheet reconnection exhibits characteristics of both
magnetic breakout for CME initiation and standard flare model. We
argue that both breakout-like and tether-cutting reconnections may
be important for CME eruptions and associated surface activities.
\end{abstract}

\keywords{Sun: corona $-$ Sun: coronal mass ejections (CMEs) $-$
Sun: flares $-$ Sun: magnetic fields}

\section{INTRODUCTION}

A number of coronal mass ejections (CMEs) showed structures
consistent with the ejection of a magnetic flux rope as it has
been reported by Chen et al. (1997), Wood et al. (1999) and Dere
et al. (1999). Therefore, magnetic flux ropes have been presumed
to be typical structures in the solar corona, and their eruptions
might be closely related to solar flares and CMEs (Forbes, 2000;
Low, 2001). A lot of studies, both analytical and numerical, tried
to explain such eruptive phenomena (Anzer, 1978; Priest, 1988;
Forbes \& Isenberg, 1991; Isenberg et al., 1993; Mikic \& Linker,
1994; Forbes \& Priest, 1995; Low, 1996; Wu et al., 1997;
Antiochos et al., 1999; Chen \& Shibata, 2000; Hu \& Liu, 2000;
Lin \& Forbes, 2000; Amari et al., 2000; Lin et al., 2001, Cheng,
et al., 2005, T\"{o}r\"{o}k \& Kliem, 2005). Most of them were
associated with a bipolar magnetic configuration and assumed that
reconnection in the current sheet below the flux rope triggers the
eruption by the so-called tether cutting of the field lines.
However, observations always show complicated magnetic
configuration and global coupling of different flux systems (see
an example described by Wang et al. (2005) and a statistic
analysis by Zhou et al. (2005)).

Two types of models are popular in the investigation of solar
eruptive phenomena: the standard flare model and the magnetic
breakout model. The standard flare model for magnetic explosions
in eruptive flares was first proposed by Sturrock (1966), and
advanced by a lot of latter studies ( Hirayama, 1974; Heyvaerts et
al., 1977; Sturrock et al., 1984; Shibata et al., 1995; Tsuneta,
1997; Shibata, 1999; Chen \& Shibata, 2000; Moore et al., 2001).
Recently, Chen \& Shibata (2000) proposed an emerging flux trigger
mechanism for CMEs, in which reconnection in the current sheet
below the rope leads to an eruption of the CME and a cusp-shaped
solar flare. All of these studies showed that a cusp structure and
a two ribbon flare occur in the lower corona, and that the
reconnection is tether-cutting at the internal current sheet.
Another type of models is the breakout model (Antiochos et
al.,1999) that involves multipolar topology and requires external
magnetic reconnection to occur on the top of the sheared arcade.
In their model the background field has a spherically symmetric
quadrupolar configuration, rather than a simple bipolar one.

Many observations have shown that CMEs and flares are often two
aspects of the same eruptive event. In a recent study (Zhang, et
al. 2005) we found that a double catastrophe exists for an
isolated flux rope embedded in a quadrupolar background field.
After the first catastrophe, the flux rope levitates in the solar
corona and two current sheets coexist with the rope, a transverse
one above and a vertical one below the rope. As a product of
interaction between the central and overlying arcades, the
transverse current sheet represents the large-scale nature of the
flux system. On the other hand, the vertical current sheet is
limited to the interior of the central arcade and comes from a
local interaction between small-scale bipoles. The coexistence of
the two current sheets differentiates the present magnetic
configuration with either the configuration of magnetic breakout
model, or that of standard flare model. In the absence of
reconnection, the flux rope may levitate in the corona in
equilibrium. The resulting magnetic configuration provides a
pre-eruption magnetic topology for a potential CME and its
associated surface magnetic activity, and meets the requirements
of magnetic breakout and standard flare models. Once reconnection
sets in across one of the two current sheets or both, an eruption
of the flux rope is inevitable, which is presumably responsible
for concurrence of CMEs and flares. To explore this possibility,
we take one of the force-free field solutions obtained by Zhang et
al. (2005) as the initial state, which is located right after the
first catastrophic point, introduce resistive dissipation in the
current sheets, and examine the dynamic evolution of the flux rope
system. The numerical results will show both breakout and
tether-cutting.

We describe the time-dependent, resistive magnetohydrodynamic
(MHD) equations and the solution procedures in section 2. We
discuss the evolution of the flux rope system in section 3, and
conclude our work in section 4.

\section{BASIC EQUATIONS AND SOLUTION PROCEDURES}

We use time-dependent resistive MHD simulations to study the
dynamic evolution of a flux rope system in the presence of
resistance. For 2.5-dimensional (2.5-D) MHD problems in spherical
coordinates (r,$\theta,\varphi$), one may introduce a magnetic
flux function $\psi(t,r,\theta)$ related to the magnetic field by
$$
{\mathbf B} = \bigtriangledown\times \left (
\frac{\psi}{r\sin\theta}\hat{\varphi} \right ) + {\mathbf
B}_\varphi, \ \ \ \ {\mathbf B}_\varphi = B_\varphi \hat{\varphi}
, \eqno (1)
$$
where $B_\varphi$ is the azimuthal component of the magnetic
field. Then the 2.5-D resistive MHD equations are cast in the
following form
$$
\frac{\partial\rho}{\partial t} + \bigtriangledown\cdot(\rho
{\mathbf v})=0, \eqno (2)
$$
$$
\frac{\partial{\mathbf v}}{\partial t} + {\mathbf
v}\cdot\bigtriangledown {\mathbf v} +
\frac{1}{\rho}\bigtriangledown p + \frac{1}{\mu\rho}
[L\psi\bigtriangledown\psi + {\mathbf
B}_\varphi\times(\bigtriangledown \times{\mathbf B}_\varphi )]
$$
$$
+\frac{1}{\mu\rho r\sin\theta}\bigtriangledown\psi
\cdot(\bigtriangledown\times{\mathbf B}_\varphi )\hat{\varphi} +
\frac{GM_{\odot}}{r^2}\hat{r}=0, \eqno (3)
$$
$$
\frac{\partial\psi}{\partial t}+ {\mathbf
v}\cdot\bigtriangledown\psi - {1 \over \mu}\eta r^{2}
\sin^{2}\theta L \psi = 0, \eqno (4)
$$
$$
\frac{\partial B_{\varphi}}{\partial t} +
r\sin\theta\bigtriangledown \cdot \left ( \frac{B_\varphi{\mathbf
v}}{r\sin\theta} \right ) + \left [ \bigtriangledown \psi\times
\bigtriangledown \left ( \frac{v_\varphi}{r\sin\theta} \right )
\right ]_\varphi-\frac{1}{r sin\theta}\bigtriangledown\eta \cdot
\bigtriangledown(\mu r \sin\theta B_{\varphi})
$$
$$
-{1\over \mu}\eta r \sin \theta L(r B_{\varphi} \sin \theta)= 0,
\eqno (5)
$$
$$
\frac{\partial T}{\partial t}+{\mathbf v}\cdot\bigtriangledown T +
(\gamma-1)T\bigtriangledown\cdot{\mathbf
v}-\frac{\gamma-1}{\rho}\eta {\mathbf j}^2 = 0, \eqno (6)
$$
where
$$
L\psi\equiv\frac{1}{r^{2}\sin^{2}\theta} \left (
\frac{\partial^{2}\psi}{\partial
r^{2}}+\frac{1}{r^{2}}\frac{\partial^{2}\psi}{\partial\theta^{2}}-
\frac{\cot\theta}{r^{2}}\frac{\partial\psi}{\partial\theta} \right
) , \eqno (7)
$$
$$
{\mathbf j} = {1\over \mu}\bigtriangledown \times {\mathbf B} =
-{1\over \mu}r\sin\theta L\psi \hat{\varphi} +{1\over
\mu}\bigtriangledown \times (B_\varphi \hat{\varphi}), \eqno (8)
$$
$\rho$ is the density, ${\mathbf v}$ is the flow velocity, $\mu$
is the vacuum magnetic permeability, $G$ is the gravitational
constant, $M_\odot$ is the mass of the Sun, $T$ is the
temperature, $\gamma$ (= 1.05) is the polytropic index, $\eta$ is
the resistivity, and ${\mathbf j}$ is the current density.

The computational domain is taken to be $1 \leq r \leq 30$ in the
unit of $R_\odot$ ($R_\odot$ is the solar radius),
$0\leq\theta\leq \pi/2$, discretized into $130 \times 90$ grid
points. The grid spacing increases according to a geometrical
series of common ratio 1.03 from 0.02 at the base ($r$ = 1) to
0.86 at the top ($r$ = 30), whereas a uniform mesh is adopted in
the $\theta$-direction. The multistep implicit scheme (Hu 1989) is
used to solve equations (2)-(6). As for the boundary conditions,
we use appropriate symmetrical conditions at the pole and equator,
and calculate the quantities at the top in terms of equivalent
extrapolation except for $B_\varphi$ and $\psi$. The magnetic
field is potential above the transverse current sheet that is
below the top boundary. Therefore, $B_\varphi$ is set to be zero
and $\psi$ is calculated from $j_\varphi$ = $-r\sin\theta L\psi$ =
0 at the top (see Hu et al., 2003; Hu, 2004; Zhang et al., 2005).

The initial corona is assumed to be isothermal and static with
$T=T_0=2\times 10^6$ K and $\rho = \rho_0 = 1.67\times 10^{-13}$
kg$\cdot$m$^{-3}$ at the coronal base, where $T_0$ and $\rho_0$
are taken to be the units for temperature and density,
respectively. Taking a characteristic value of 0.01 for $\beta$,
the ratio of gas pressure to magnetic pressure, leads to a
characteristic value of $\psi_0$ = $(2\mu \rho_0 R T_0 R_\odot^4 /
\beta )^{1/2}$ = 5.69$\times 10^{14}$ Wb, taken to be the unit of
$\psi$. Other units of interest are $B_0$ = $\psi_0/R_\odot^2$ =
1.18$\times 10^{-3}$ T for field strength, $v_A$ =
$B_0/(\mu\rho_0)^{1/2}$ = 2570 km$\cdot$s$^{-1}$ for velocity,
$\tau_A$ = $R_\cdot /v_A$ = 271 s for time, and $j_0 = B_0/(\mu
R_\odot )$ = 1.35$\times 10^{-6}$ A$\cdot$m$^{-2}$ for electric
current density.

We choose a force-free field solution as the initial magnetic
field. This solution was obtained by Zhang et al. (2005) right
after the first catastrophic point, characterized by an isolated
flux rope levitating in the corona and accompanied by two current
sheets, a transverse one above and a vertical one below the rope.
The annular magnetic flux per radian is 0.6 in the unit of
$\psi_0$, and the axial magnetic flux is 0.0416 in the unit of
$\psi_0$ for the flux rope, and both of them are conserved during
subsequent dynamic evolutions of the flux rope system. The
magnetic energy of the initial field is 1.71, which is still
larger than the energy of the associated partially open field,
1.662, by 2.9\% (see Zhang, et al.,2005). The excess energy is
obviously in favor of high-speed CMEs.

The initial field chosen above is in equilibrium in the ideal MHD
regime, but will certainly evolve into a dynamic state once
reconnection sets in across the current sheets. The temporal
evolution of the whole system depends on how reconnection occurs
in the two current sheets. Three cases will be treated, labelled
A, B, and C hereinafter, and they differ in the sequence of
reconnection. Reconnection starts simultaneously in the two
current sheets in case A, first in the transverse current sheet
and later on in the vertical one in case B, and in the opposite
order in case C. To control the sequence of reconnection, we
introduce a critical current density for each current sheet,
denoted by $j_t$ for the transverse current sheet and $j_v$ for
the vertical one. When the current density nearby the transverse
current sheet exceeds $j_t$ or that nearby the vertical current
sheet exceeds $j_v$, the resistivity of $\eta$ is set to be 0.01,
and $\eta$ is set to be 0 elsewhere. Consequently, we may simply
set $j_t$ larger than the initial peak current density in the
transverse current sheet to delay reconnection or smaller than the
initial peak current density to start reconnection across the
sheet. Notice that a larger value of $j_t$ just causes a delay of
reconnection, rather than prohibits it. As a mater of fact, the
current density in the transverse current sheet grows with time
during the rope eruption, so it may eventually exceed $j_t$
somewhere, leading to a delayed onset of reconnection in the
sheet. The same is the case for the vertical current sheet. Such
an expedient measure is somewhat artificial but satisfies our
purpose. Through tentative calculations, we find that the initial
peak current density is 5.3 in the transverse current sheet and
22.1 in the vertical current sheet. Consequently, we choose
($j_t$, $j_v$) = (5, 20) for case A, (5, 40) for case B, and (10,
20) for case C.

\section{SIMULATION RESULTS}

As mentioned in the previous section, we intend to discuss three
cases, a simultaneous reconnection in the transverse and vertical
current sheets for case A, a first reconnection in the transverse
current sheet followed by a second in the vertical current sheet
for case B, and a first reconnection in the vertical current sheet
followed by a second in the transverse current sheet for case C.
In each case, we use the height of the rope axis relative to the
solar surface, $h_a$, to mark the position of the flux rope. For
the initial state, we have $h_a$ = 1.70.

In case A, reconnection occurs simultaneously in the transverse
and vertical current sheets. Figures 1a-1c show the magnetic
configuration at three separate times, along with the temperature
distribution in color. Figure 1a corresponds to the initial state,
and resistive dissipation is switched on in both current sheets at
$t$ = 0. Since then, high temperature appears in the current sheet
regions because of reconnection, and the flux rope erupts upward,
as shown in Figures 1b and 1c. The rope is immediately accelerated
without an initial slow rising phase as shown in Figure 2 (solid),
and it gains its maximum eruption speed of 595 km$\cdot$s$^{-1}$
at about $t$ = 5 $\tau_A$, when $h_a$ reaches 2.31 (Figure 1c).
Meanwhile, a cusp-shaped structure with high temperature is
clearly seen in Figure 1c, a typical feature of flares. Also, a
high temperature structure appears in the corona right above the
cusp structure at 1.5 in height.

\placefigure{fig1}

\placefigure{fig2}

In case B, reconnection occurs first in the transverse current
sheet, and then with the growth of the current in the vertical
current sheet, reconnection follows over there. Figures 3a-3c show
the magnetic configuration and temperature distribution at several
separate times. At $t$ = 1 $\tau_A$ when reconnection is initiated
in the transverse current sheet, the temperature along the sheet
rises. As shown by dashed line in Figure 2, the flux rope's speed
increases with time very slowly until reconnection sets in across
the vertical current sheet at about $t$ = 7 $\tau_A$ (Figure 3b).
Then the flux rope undergoes a slight deceleration of short
duration (about 1 $\tau_A$), followed by a quick acceleration. The
rope gains its maximum speed of 670 km$\cdot$s$^{-1}$ at about $t$
= 12 $\tau_A$, when $h_a$ reaches 2.34 (Figure 3c). Similarly, a
cusp-shaped structure with high temperature and a coronal high
temperature structure also appear in this case.

\placefigure{fig3}

In case C, reconnection occurs first in the vertical current
sheet, and then with the growth of the current in the transverse
current sheet, reconnection follows over there. Figures 4a-4c show
the magnetic configuration and temperature distribution at several
separate times. At $t$ = 1 $\tau_A$ when reconnection occurs only
in the vertical current sheet, the temperature along the sheet
rises, as shown in Figure 4a. It can be seen from the dash-dotted
profile in Figure 2 that the flux rope is accelerated before that
time, slightly decelerated afterwards about 2 $\tau_A$ in
duration, and then accelerated again with a much larger
acceleration. The flux rope gains a maximum speed of 568
km$\cdot$s$^{-1}$ at about $t$ = 10 $\tau_A$, when $h_a$ reaches
3.0 (Figure 4c). This case differs from case B in that the
cusp-shaped structure is formed much earlier: it becomes clear as
early as $t$ = 4.7 $\tau_A$ (Figure 4b). And at that time the
reconnection initiates in the transverse current sheet.

\placefigure{fig4}

In summary, magnetic reconnection causes an eruption of the flux
rope and the formation of a cusp-shaped structure of high
temperature in all three cases. The former is presumably a
manifestation of CMEs whereas the latter characterizes a
two-ribbon flare. The reconnection sequence plays a critical role
in the motion of the erupting flux rope and the formation of the
cusp-shaped structure. The reconnection in the transverse current
sheet is apt to produce a gradual acceleration of the flux rope
but a higher peak speed and has little bearing on the formation of
the cusp-shaped structure. On the other hand, the reconnection in
the vertical current sheet is directly responsible for the
formation of the cusp-shaped structure and leads to an immediate
acceleration of the flux rope. It is interesting to note that a
short term deceleration occurs before the rapid acceleration
caused by reconnection across the vertical current sheet, as seen
in cases B and C. Presently we do not know exactly why the flux
rope has such a behavior. A possible reason might be that the
magnetic pressure decreases right beneath the flux rope when
reconnection starts in the vertical current sheet. High resolution
observations at both optical and radio bands show indications that
flux systems shrink first during the impulsive phases of flares,
and then explode later in the main phases of flares (Ji et al.,
2004, Li \& Gan, 2005). This seems to be consistent with the
simulation results of reconnection occurring in the vertical
current sheet in cases B and C. More careful work needs to be done
in order to judge whether this is a common behavior of flux rope
dynamics in the flare impulsive phase. Incidentally, since we have
not considered the background solar wind, the flux rope's speed
decreases after they obtain a peak speed in all three cases.

\section{Concluding Remarks}

Using time-dependent resistive MHD simulations, we find solutions
associated with an isolated coronal flux rope embedded in a
quadrupolar background field and accompanied by a transverse
current sheet above and a vertical current sheet below the rope.
Reconnection may occur in the current sheets either simultaneously
or one after another. The present model agrees with the breakout
model (Antiochos, 1999; Lynch, et al. 2004) if reconnection is
initiated in the transversal current sheet, and it returns to the
standard flare model (Chen \& Shibata, 2000) if reconnection is
initiated in the vertical current sheet. Nevertheless, we argue
that both breakout-like external reconnections and tether-cutting
internal reconnections are essential to the magnetic eruption in
general. Williams et al. (2005) showed observational evidence for
the presence of both tether-cutting and breakout in eruptive
events. Our simulations just combine the two models together,
which is probably more relevant to observations that many eruptive
events occur in background fields of quadrupolar magnetic
configuration (Sterling \& Moore, 2004; Sterling \& Moore, 2004;
Gary \& Moore, 2004).

The present magnetic configuration and the dynamical evolution
shed new light on understanding the relationship between CMEs and
flares, which is a topic with great interest and hot debates. More
and more investigations prefer a closer and rather intrinsic
association between CMEs and surface activities (see Zhang et
al.a,b; Zhou et al. 2003).

Zhang et al. (2001a) reported that the kinematic evolution of CMEs
can be described in a three-phase scenario: the initiation phase,
the impulsive acceleration phase, and the propagation phase.
Furthermore, they found that following the initiation phase, the
CME displays an impulsive acceleration phase, which starts almost
simultaneously with the flare onset time. After the acceleration
phase the CME undergoes a propagation phase. And Zhang et al.
(2001b) found a halo CME that moved slowly in the initial phase,
and was later on accelerated and erupted. This is consistent with
our case B, in which reconnection starts first in the transversal
current sheet, leading to a slow upward motion of the CME, and
subsequently, because of reconnection onset in the vertical
current sheet, the CME acceleration is quickened until it reaches
the maximum speed. In other words, the breakout first occurs and
the tether-cutting follows. However, this is just one possibility,
the other two cases we work out would appear in different
circumstances. Zhou et al.(2003) gave a statistic result that
$59\%$ of the selected 197 halo CMEs initiate earlier than the
flare onset and $41\%$ are preceded by flare onsets. The latter
samples may relate to our case C. Furthermore, Zhang et al.
(2001a) also found one CME that did not show an initiation phase,
but was immediately accelerated to the maximum speed. This example
is very similar to our case A in which reconnection occurs
simultaneously in the two current sheets.

Another point is worthy of mentioning as to the effect of the
reconnection sequence on the maximum speed of CMEs. The flux rope,
identified as the CME here, has the largest speed when
reconnection starts first in the transverse current sheet. On the
other hand, the maximum speed is the lowest when reconnection
starts first in the vertical current sheet. This implies that the
reconnection sequence may affect the maximum speed of CMEs.

\acknowledgments

The authors are greatly indebted to the anonymous referee for
helpful comments and valuable suggestions on the manuscript. One
of the authors (YZZ) thanks J.Y. Ding for kind assistance in
coding and P.F. Chen for helpful discussions. The work is
supported by the National Natural Science Foundation of China
(10233050, 40274049) and the National Key Basic Science Foundation
(TG2000078404).

\clearpage

\begin{figure}
\figurenum{1} \epsscale{0.40} \figcaption[fig1.jpg]{Magnetic
configuration in black solid curves and temperature distribution
in color at several separate times for case A, in which
reconnection occurs simultaneously in the transverse and vertical
current sheets. The cool blue and the hot red correspond to $2.51
\times 10^6$ K and $1.0\times 10^7$ K, respectively. \label{fig1}}
\end{figure}

\begin{figure}
\figurenum{2} \epsscale{0.60} \figcaption[fig2.jpg]{The velocity
of the flux rope axis versus time for each of the three cases A
(solid), B (dashed), and C (dash-dotted). \label{fig2} }
\end{figure}

\begin{figure}
\figurenum{3}\epsscale{0.40}\figcaption[fig3.jpg]{Same as Fig. 1
but for case B, in which reconnection occurs first in the
transverse current sheet and then in the vertical one.
\label{fig3}}
\end{figure}

\begin{figure}
\figurenum{4} \epsscale{0.40} \figcaption[fig4.jpg]{Same as Fig. 3
but for case C, in which reconnection occurs first in the vertical
current sheet and then in the transverse one.  \label{fig4}}
\end{figure}

\clearpage

\end{document}